# Infrared Spectroscopy with Visible Light

*Dmitry A. Kalashnikov[1], Anna V. Paterova[1], Sergei P. Kulik[2], and Leonid A. Krivitsky[1]*

[1]Data Storage Institute, Agency for Science, Technology and Research (A*STAR), 117608 Singapore

[2] Department of Physics, M. V. Lomonosov Moscow State University, 119991 Moscow, Russia

**Spectral measurements in the infrared (IR) optical range provide unique fingerprints of materials which are useful for material analysis, environmental sensing, and health diagnostics[1]. Current IR spectroscopy techniques require the use of optical equipment suited for operation in the IR range, which faces challenges of inferior performance and high cost. Here we develop a spectroscopy technique, which allows spectral measurements in the IR range using visible spectral range components. The technique is based on nonlinear interference of infrared and visible photons, produced via Spontaneous Parametric Down Conversion (SPDC)[2,3]. The intensity interference pattern for a visible photon depends on the phase of an IR photon, which travels through the media. This allows determining properties of the media in the IR range from the measurements of visible photons. The technique can substitute and/or complement conventional IR spectroscopy techniques, as it uses well-developed optical components for the visible range.**

Entangled photons continue to play a crucial role in advancing many areas of quantum technologies, including cryptography[4,5], computing[6,7] and metrology[8,9]. They can be obtained using a variety of methods, with SPDC in non-linear optical crystals, being well established[10]. We consider a specific type of interference technique, referred to as a non-linear interferometry, which is analogue to a conventional Mach-Zehnder or Young interferometer but with the two splitting mirrors being substituted with two SPDC crystals[3]. In a non-linear interferometer, two SPDC crystals are pumped by a common laser, so that down-converted photons (signal and idler) from one crystal are injected into the second crystal. Signal and idler photons from the two crystals interfere and produce a distinctive interference pattern in frequency and spatial domains. Depending on experimental configuration one can observe interference either in intensity or in the second-order correlation function[11-13].

One remarkable feature of non-linear interferometers is that the interference pattern for signal photons is determined by a total phase, acquired by all three propagating photons: the signal, the idler and the pump[2,3]. This is different than conventional interferometry, where the interference pattern is defined solely by the phase of the signal photon. From the interference pattern of the signal photon, it is possible to infer a relative phase of an idler photon. Actual detection of idler photons is not required. This scheme has found its applications in imaging with undetected photons[14], interferometry below the shot-noise[15] and spectroscopy of Raman scattering by polaritons[16].

Here we address the problem of optical investigation of materials in the infrared (IR) range. IR spectroscopy is a powerful tool for chemical analysis and sensing[1]. A number of well-developed techniques are commercially available, including transmission spectroscopy and Fourier Transform IR (FT-IR) spectroscopy. The common requirement for these techniques is the use of IR light sources, photodetectors, and optical elements. These components often have inferior performance and higher cost in comparison to their visible range counterparts. Here, for the first time, the nonlinear interferometer is applied to simultaneous measurements of the refractive index and absorption of a medium in the broad IR range without the need of IR equipment. The IR properties of the medium are inferred from the interference pattern for the signal photons in the visible range, measured with accessible optical instruments.

A nonlinear Mach-Zehnder interferometer comprised of two SPDC crystals pumped by a common laser is the foundation of this technique, see Fig.1. Two identical crystals with thickness $L$ are separated by distance $L_m$. The crystals are set to produce signal and idler photons in the visible and IR range, respectively. We assume a quasi-collinear regime of SPDC, when the diameter of the pump beam $a \gg L_m * \max\{\theta_s, \theta_i\}$, where $\theta_{s,i}$ is the emission angle for signal (*s*) and idler (*i*) photons. In this case the dependence of the intensity of signal photons on the wavelength $\lambda_s$ and the emission angle $\theta_s$ is given by[17]:

$$I_s(\lambda_s, \theta_s) \propto \frac{1}{2}\left[sinc\left(\frac{\delta}{2}\right)\right]^2 \{1 + cos(\delta + \delta^m)\}, \tag{1}$$

$$\delta(\lambda_s, \theta_s) = (k_p - k_s - k_i)L, \qquad \delta^m(\lambda_s, \theta_s) = (k_p^m - k_s^m - k_i^m)L_m,$$

where $k_j$ and $k_j^m$ are the wavevectors in the SPDC crystal and in the gap, respectively, and $j = s, i, p$ for signal, idler and pump photons. The wavevectors are given by $k_j = 2\pi n_j/\lambda_j$, and $k_j^m = 2\pi n_j^m/\lambda_j$, where $n_j$ ($n_j^m$) is the refractive index of the SPDC crystal (of the gap), and $\lambda_j$ is a wavelength. We consider that SPDC crystals are transparent for signal, idler and pump photons, and corresponding refractive indices $n_s, n_i, n_p$ are known. In equation (1) the first term defines the SPDC spectrum of an individual crystal, and the second term defines modulation due to interference. The modulation term depends on the total phase acquired by signal, idler and pump photons in the crystals and in the medium.

When the medium with absorption at the wavelength of idler photons is introduced in the gap between the crystals, the interference pattern is given by[17]:

$$I_s(\lambda_s, \theta_s) \propto \frac{1}{2}\left[sinc\left(\frac{\delta}{2}\right)\right]^2 \{1 + |\tau_i^m|cos(\delta + \delta^m)\} \tag{2}$$

where $\tau_i^m$ is the transmission coefficient of the medium for the idler photon. We assume $|\tau_i^m| \propto exp(-\alpha_i^m L_m)$, where $\alpha_i^m$ is the Bouguer absorption coefficient. Visibility of the

interference for signal photons, defined as $V_s = \frac{I_{s,max} - I_{s,min}}{I_{s,max} + I_{s,min}}$, vanishes with $\tau_i^m \to 0$. It can be used for estimation of the absorption coefficient $\alpha_i^m = -\frac{\ln(V_s)}{L_m}$.

From equation (2) it follows that introduction of the medium between the crystals shifts the interference fringes due to an additional phase, proportional to the refractive index $n_j^m$. Also the visibility of the fringes decreases depending on $\alpha_i^m$. Without loss of generality we assume that the medium is transparent for signal and pump photons, and that the refractive indices for signal and idler photons $n_{s,p}^m$ are known. Then, by fitting the measured interference pattern with equation (2), we can infer both the refractive index $n_i^m$ and the absorption coefficient $\alpha_i^m$ for the medium at the wavelength of idler photons. Note that $n_{s,p}^m$, can be measured using accessible visible light equipment.

Carbon dioxide gas ($CO_2$) has been chosen as the medium of interest in our experiments due to its importance as a greenhouse gas and also for its use in clinical breath diagnostics. $CO_2$ is transparent in the visible and it absorbs light in the mid-IR, - the wavelength range where we demonstrate the applicability of our technique.

We studied the absorption line of $CO_2$ associated with an asymmetric stretching mode at the wavelength of 4.3 micron. The wavelength of an idler photon in the SPDC coincides with the center of the absorption line of $CO_2$. The wavelength of the signal photon is calculated according to the energy conservation law for the SPDC[10]. For the case of the absorption line at 4.3 micron, the wavelength of the signal photon is at 608 nm.

Interference patterns of the intensity in angular-wavelength coordinates for the signal photon at zero concentration of $CO_2$ and at $CO_2$ pressure of 7.7 Torr are shown in Fig.2a and Fig. 2b, respectively. The insets in each figure correspond to an angular distribution of the intensity of signal photons at the selected wavelength, denoted by a vertical yellow line. Interference fringes clearly shift and their visibility is reduced as a result of the interaction of idler photons with $CO_2$, see the animation in the supplementary materials section.

First, we obtain the dependence of the absorption coefficient and the refractive index on the wavelength of the idler photon. Angular cross-sections at different wavelengths of signal photons are fitted by equation (2) with $n_i^m, \alpha_i^m$ being the only fitting parameters (Levenberg–Marquardt algorithm, confidence level ~ 95%). Values of the refractive index of $CO_2$ in the visible range are taken from the literature[18], and we take their dependence on pressure to be linear.

The absorption coefficient $\alpha_i^m$ is calculated from the visibility of the interference pattern. The reference value of the visibility is determined from a measurement with vacuum (20 mTorr) between the crystals. Experimental results at a $CO_2$ pressure of 10.5 Torr are shown in Fig. 3a. The dependence exhibits a peak at 4.3 μm with the full width at a half maximum of 140 nm. The result is in good agreement with calculations based on data from Hitran[19], assuming the spectral resolution of our setup.

The dependence of the refractive index on the wavelength at the same pressure of $CO_2$ is calculated from the relative shift of the interference fringes from the vacuum case. The experimental dependence is shown in Fig. 3b and it is found in good agreement with the theory based on Kramers-Kronig equations, applied to the absorption data.

We also study the dependence of the absorption coefficient $\alpha_i^m$ as a function of gas pressure at different wavelengths of the idler photon, see Fig. 4a. Steeper dependences correspond to wavelengths closer to the absorption line. The dependence of the refractive index $n_i$ on the pressure for different wavelengths are shown in Fig. 4b.

The obtained dependences exhibit some nonlinearity in the vicinity of the absorption line. We refer this fact to different mechanism of broadening of the absorption line at different pressures[20,21]. At lower pressure the line exhibits Doppler broadening due to motion of molecules. At higher pressure the line takes the Loretzian shape due to collision broadening. There are intermediate points between these two cases when both mechanisms play a role. However, away from the absorption line the measurements of refractive index (green triangles at Fig. 4b) show linear dependence on pressure, which yields *n=1* at *P=0* Torr.

In conclusion, the suggested approach allows direct measurement of real and imaginary parts of a complex refractive index of a medium in broadband IR-range by detecting photons in the visible range. The technique relies on non-linear interference of frequency entangled photons, produced via SPDC. The method does not suffer from losses in the optical path due to absorption by water vapor, which is a known issue for conventional IR spectroscopy. . The accuracy of the technique reaches $\sim 5*10^{-6}$ in determination of the refractive index and $\sim 10^{-3}$ cm$^{-1}$ in determination of the absorption coefficient, which is comparable with state-of-art IR methods[1].

Due to broad tunability of the SPDC source, this method can be tailored to satisfy requirements for a desired operation range and spectral resolution, including the far-IR and terahertz ranges[22,23]. At the same time further improvement to the sensitivity is feasible by using interferometers with larger base and/or multi-pass configurations. This developed approach helps to overcome some limitations of conventional IR-spectroscopy techniques and represent a practical alternative or complementary technique for applications in material analysis and sensing.

## Methods

**Optical setup.** The pump beam from a 532 nm CW laser (Spectra-Physics, Millennia V) with the waist *a*=2 mm is sent through two identically oriented MgO:LiNbO$_3$ crystals (Dayoptics, doped with 5% Mg, transparency range from 0.45 to 5 µm) with thickness *L*=0.5 mm. Optical axes of the crystals are cut at 50 degrees to the surface for type-I SPDC (*e→oo* interaction). The crystals are placed in a custom made vacuum chamber (with two optical windows) at a distance of $L_m$=25 mm from each other. Changing the orientation of the crystal optical axis allows tuning of the frequency of signal and idler photons and to study $CO_2$ resonances at different wavelengths. The chamber is pumped down to 20 mTorr and vacuum is used as a reference media with $n_{i,s,p}$=1. The $CO_2$ gas (National Oxygen, purity ≥ 99.99%) is fed into the vacuum chamber from a gas cylinder, and its pressure is controlled by a gauge sensor (Granville-Phillips, 275 Mini-Convectron). After passing through the crystals the pump beam is filtered by two notch filters (Semrock, NF03-532E-25). The SPDC radiation is imaged by a lens with *f*=500 mm onto the input slit of a spectrometer (Acton, SpectraPro 2300i). The slit of the spectrometer is positioned at the lens focal plane. At the output of the spectrometer the image of the slit, illuminated by the SPDC, is formed providing a two-dimensional wavelength-angular distribution, which is detected by an electronically multiplied CCD camera (Andor, iXon3 888, detection range 300-1000 nm, pixel area 13x13 μm$^2$). Each CCD pixel in the horizontal scale is assigned to the corresponding wavelength, and each pixel in vertical scale is assigned to the corresponding angle. Each image is saved to the PC and then processed by Mathematica and Origin software.

**Dependence of the refractive indexes on pressure.** We use values of the refractive index for $CO_2$ in the visible range from the literature[18], and assume that the dependence of the refractive index $n$ on the gas pressure $P$ follows: $n(P) = 1 + \frac{P(n_0-1)}{P_0(1+(T-T_0)/T_0)}$, where $n_0$ is a refractive index at the atmospheric pressure $P_0$, $T$ is the temperature of the gas (300 K for our case), and $T_0$=273K.

## References

1) Stuart, B. H. *Infrared Spectroscopy: Fundamentals and Applications*, (Wiley Interscience, 2004).

2) Zou, X. Y., Wang, L. J. & Mandel, L. Induced coherence and indistinguishability in optical interference, *Phys. Rev. Lett.* **67**, 318 (1991).

3) Mandel, L. Quantum effects in one-photon and two-photon interference, *Rev. Mod. Phys.* **71**, S274 (1999).

4) Gisin, N., Ribordy, G, Tittel, W. & Zbinden, H. Quantum Cryptography, *Rev. Mod. Phys.* **74**, 145-195 (2002).

5) Gisin, N. & Thew, R. Quantum communication, *Nature Photon.* **1**, 165-171 (2007).

6) Knill, E., Laflamme, R. & Milburn, G. J. A scheme for efficient quantum computation with linear optics, *Nature* **409**, 46-52 (2001).

7) Ladd, T. D., Jelezko, F., Laflamme, R., Nakamura, Y., Monroe, C. & O'Brien, J. L. Quantum computers, *Nature* **464**, 45-53 (2010).

8) Giovannetti, V., Lloyd, S. & Maccone, L. Advances in quantum metrology, *Nature Photon.* **5**, 222-229 (2013).

9) Abadie, J. *et al.* A gravitational wave observatory operating beyond the quantum shot-noise limit, *Nature Phys.* **7**, 962-965 (2011).

10) Klyshko, D. N. *Photon and Nonlinear Optics*, (Gordon and Breach Science, 1988).

11) Burlakov, A. V., Chekhova, M. V., Klyshko, D. N., Kulik, S. P., Penin, A. N., Shih, Y. H. & Strekalov, D. V. Interference effects in spontaneous two-photon parametric scattering from two macroscopic regions, *Phys. Rev. A* **56**, 3214 (1997).

12) Korystov, D. Yu., Kulik, S. P. & Penin, A. N. Rozhdestvenski hooks in two-photon parametric light scattering, *JETP Lett*. **73**, 214-218 (2001).

13) Kulik, S. P., Maslennikov, G. A., Merkulova, S. P., Penin, A. N., Radchenko, L. K. & Krashenninikov, V. N. Two-photon interference in the presence of absorption, *Journal of Experimental and Theoretical Physics* **98**, 31-38 (2004).

14) Lemos, G. B., Borish, V., Cole, G. D., Ramelow, S., Lapkiewicz, R. & Zeilinger, A. Quantum imaging with undetected photons, *Nature* **512**, 409–412 (2014).

15) Hudelist, F., Kong J., Liu, C., Jing, J., Ou, Z. Y. & Zhang, W. Quantum metrology with parametric amplifier-based photon correlation interferometers, *Nat. Commun*. **5**, 3049 (2014).

16) Polivanov, Yu. N. Raman scattering of light by polariton, *Sov. Phys. Usp.* **21**, 805–831 (1978)

17) Klyshko, D. N. Ramsey interference in two-photon parametric scattering, *JETP* **77**, 222-226 (1993).

18) Bideau-Mehu, A., Guern, Y., Abjean, R. & Johannin-Gilles, A. Interferometric determination of the refractive index of carbon dioxide in ultraviolet region. *Opt. Commun*. **9**, 432-434 (1973)

19) http://hitran.iao.ru

20) Heineken, F. W. & Battaglia, A. Absorption and refraction of ammonia as a function of pressure at 6 mm wavelength, *Physica* **XXIV**, 589-603 (1958).

21) Burch, D. E. & Williams, D. Total absorptance by nitrous oxide bands in the infrared, *Applied Optics* **1**, 473-482 (1962).

22) Tonouchi, M. Cutting-edge terahertz technology, *Nature Photon.* **1**, 97-105 (2007).

23) Kailash, C. J., Covert, P. A., & Hore, D. K. Phase measurement in nondegenerate three-wave mixing spectroscopy, *J. Chem. Phys*.**134**, 044712 (2011).

## Acknowledgements

This work is supported by DSI core funds within the framework of Quantum Sensors program. We would like to thank Guillaume Vienne, Reuben Bakker, Gleb Maslennikov and Dmitriy Kupriyanov for stimulating discussions and advice on the experiment.

## Author Contributions

D.A.K. and L.A.K. assembled the experimental setup and conducted the measurements. A.V.P. analyzed the data and carried out numerical simulations. L.A.K. and S.P.K. conceived the idea and designed the experiment. All the authors contributed to preparation of the manuscript.

## Competing financial interests

The authors declare no competing financial interests.

## Correspondence and requests for materials should be addressed to L.A.K.

e-mail: Leonid_Krivitskiy@dsi.a-star.edu.sg.

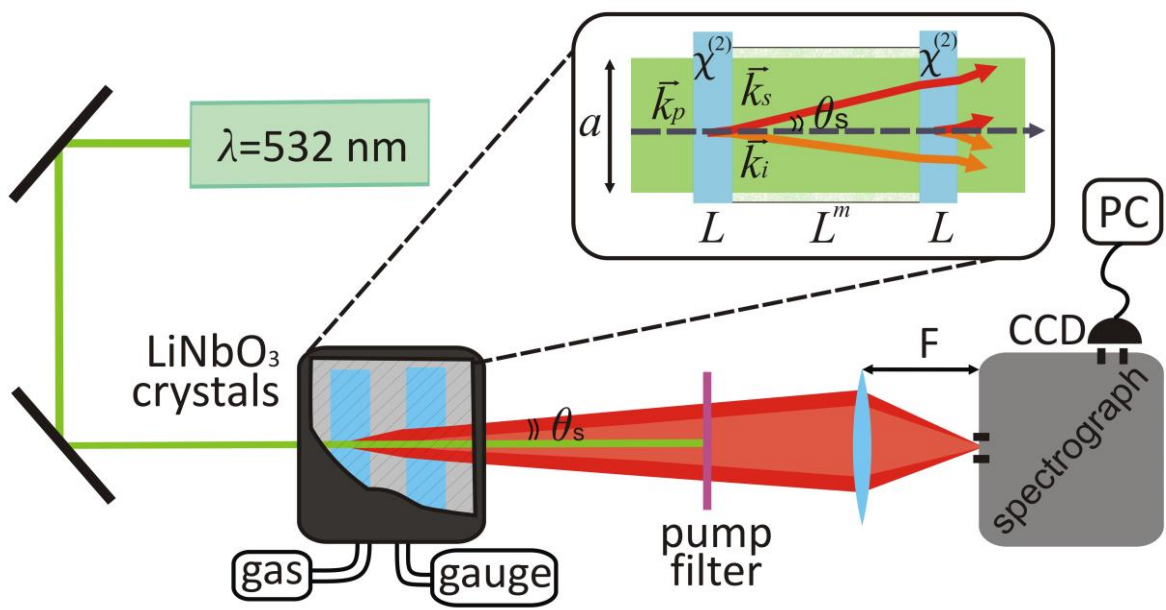

**Figure 1 | The experimental setup.** A CW-laser at 532 nm pumps two nonlinear crystals where SPDC occurs. The crystals are placed in a vacuum chamber. $CO_2$ is injected into the chamber. The interference pattern of the SPDC from the two crystals is imaged by a lens onto a slit of a spectrograph and recorded by a CCD camera.

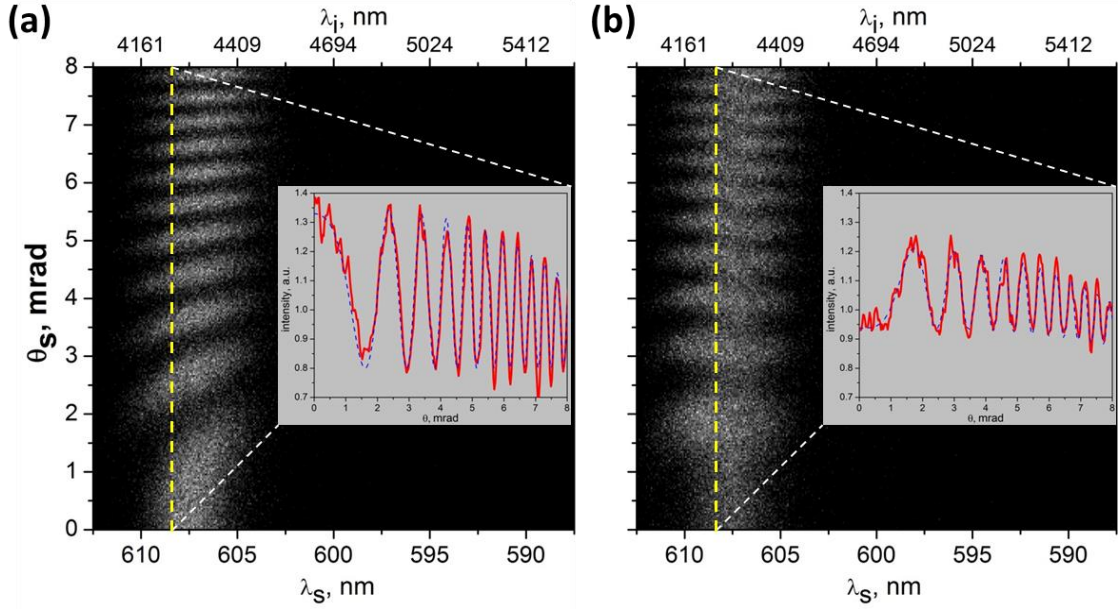

**Figure 2 | The angular-wavelength intensity distribution for signal photons from two SPDC crystals. a, b,** The intensity distribution obtained at the pressure of 0 Torr (vacuum) (**a**) and at the pressure of 7.7 Torr for $CO_2$ gas (**b**). Absorption of idler photons by the $CO_2$ leads to the shift of the fringes and the decrease of their visibility. The top axis shows corresponding wavelengths of idler photons, which are set to be close to the $CO_2$ resonance. Insets show angular distributions of the intensity at the wavelength of signal photons 608,5 nm (idler wavelength 4230 nm) denoted by a yellow dotted line. In the insets red lines show experimental data and blue dashed lines show numerical fit of the experimental data.

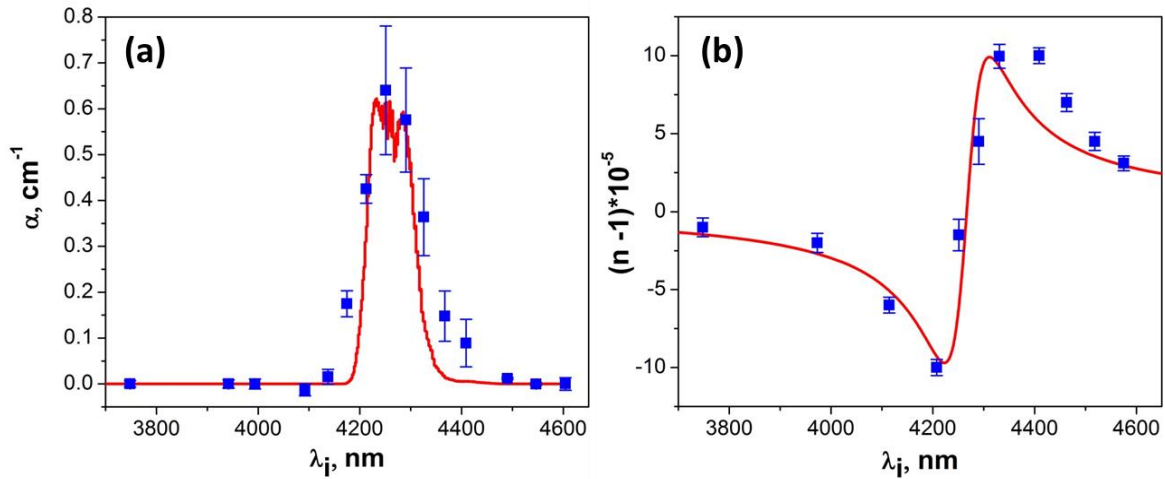

**Figure 3 | The dependence on wavelength. a, b,** The dependence of the absorption coefficient (**a**) and the refractive index (**b**) in the vicinity of the resonance for $CO_2$ at the pressure of 10.5 Torr. Blue squares denote experimental and red line denotes theoretical calculations using Hitran data (**a**) and a Kramers-Kronig relation (**b**). The error bars show ±st.dev. They are estimated within the fitting procedure with equation (2) assuming normal distribution of the intensity in the interference pattern.

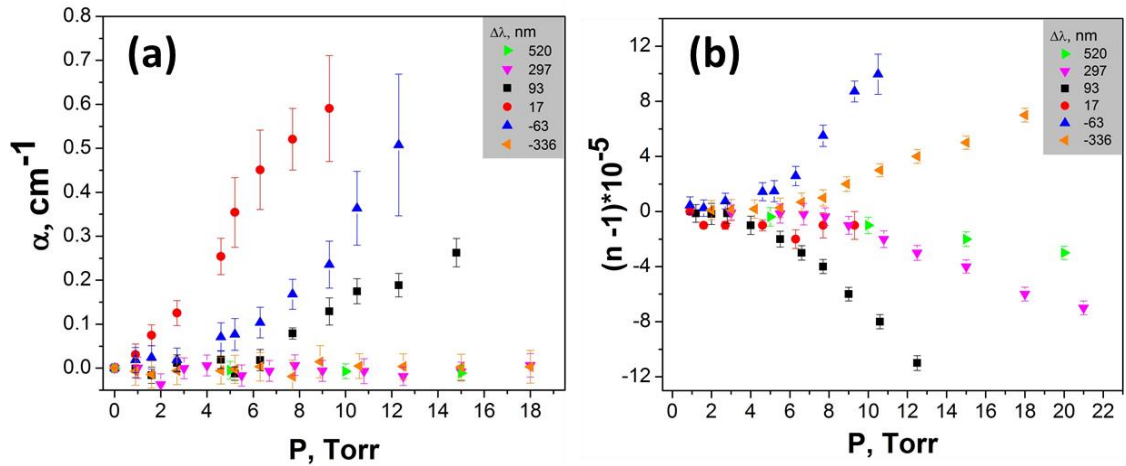

**Figure 4 | The dependence on pressure. a, b,** Dependences of the absorption coefficient (**a**) and the refractive index (**b**) on the pressure of $CO_2$ at various wavelengths. In the inset the detuning is indicated from the central wavelength of absorption line. The error bars show ±st.dev. They are estimated within the fitting procedure with equation (2) assuming normal distribution of intensity in interference pattern.